\documentclass[a4paper,11pt]{article}
\pdfoutput=1 

\usepackage{jheppub} 
\usepackage[T1]{fontenc} 
\usepackage{xcolor}
\usepackage{subfigure}
\usepackage[export]{adjustbox}
\usepackage{url}
\usepackage[utf8]{inputenc}
\usepackage[english]{babel}

\usepackage{appendix}

\newtheorem{theorem}{Theorem}

\newcommand{\ket}[1]{\left|#1\right\rangle}
\newcommand{\bra}[1]{\left\langle #1\right|}
\newcommand{\norm}[1]{\left\lVert #1\right\rVert}

\newcommand{\rhoqft}{{\boldsymbol{\rho}}}
\newcommand{\sigmaqft}{{\boldsymbol{\sigma}}}

\title{\boldmath Entanglement island, miracle operators and the firewall}

\author{Xiao-Liang Qi}

\affiliation{Stanford Institute for Theoretical Physics, Stanford University, Stanford, California 94305, USA}

\abstract{In this paper, we obtain some general results on information retrieval from the black hole interior, based on the recent progress on quantum extremal surface formula and entanglement island. We study an AdS black hole coupled to a bath with generic dynamics, and ask whether it is possible to retrieve information about a small perturbation in the interior from the bath system. We show that the one-norm distance between two reduced states in a bath region $A$ is equal to the same quantity in the bulk quantum field theory for region $AI$ where $I$ is the entanglement island of $A$. This is a straightforward generalization of bulk-boundary correspondence in AdS/CFT. However, we show that a contradiction arises if we apply this result to a special situation when the bath dynamics includes a unitary operation that carries a particular measurement to a region $A$ and send the result to another region $W$. Physically, the contradiction arises between transferability of classical information during the measurement, and non-transferability of quantum information which determines the entanglement island. 

We propose that the resolution of the contradiction is to realize that the state reconstruction formula does not apply to the special situation involving interior-information-retrieving measurements. This implies that the assumption of smooth replica AdS geometry with boundary condition set by the flat space bath has to break down when the particular measurement operator is applied to the bath. Using replica trick, we introduce an explicitly construction of such operator, which we name as ``miracle operators''. From this construction we see that the smooth replica geometry assumption breaks down because we have to introduce extra replica wormholes connecting with the ``simulated blackholes'' introduced by the miracle operator. We study the implication of miracle operators in understanding the firewall paradox. }

\begin{document} 
\maketitle
\flushbottom

\section{Introduction}

Since the discovery of Hawking radiation\cite{hawking1975particle}, the black hole information paradox remains an important open question in physics. The key question in black hole information paradox is the fate of information carried by objects falling into a black hole. If unitarity of quantum mechanics is preserved in the presence of gravity, one expects the Hawking radiation cannot be always thermal, and the entropy of an evaporating black hole has to decrease at late time, known as the Page curve.\cite{page1993average} An important recent progress towards solving the black hole information paradox is the discovery of entanglement island\cite{penington2020entanglement,almheiri2019entropy}. The key idea of the entanglement island works is to generalize the Hubeny-Rangamani-Takayanagi (HRT) formula\cite{ryu2006holographic,hubeny2007covariant} with quantum corrections\cite{faulkner2013quantum} to the entanglement entropy of a region in the radiation. In standard holographic duality, the entanglement entropy of a boundary region $A$ is dual to the area of an extremal surface $\gamma_A$ which is homologous to $A$, with the location of $\gamma_A$ determined by extremizing the generalized entropy $\frac{\left|\gamma_A\right|}{4G_N}+S_{bulk}(\Sigma_A)$. Here $\Sigma_A$ is the region in the bulk with $\gamma_A$ and $A$ as its boundary, which is known as the entanglement wedge of $A$. Ref. \cite{penington2020entanglement,almheiri2019entropy} and more recent works have generalized this formula to a coupled system of a holographic conformal field theory (CFT) and a bath system. $S_{bulk}(\Sigma_A)$ is generalized to the entanglement entropy of a region $\Sigma_A$ in the quantum field theory with fixed geometry background, where $\Sigma_A$ includes the $A$ region itself and possibly an additional region $I$ in the holographic bulk. Region $I$ is determined by the same extremization procedure, which is called an ``entanglement island'' when it is nontrivial. The entanglement island formula can be derived using a replica calculation, where different replicas are connected by a replica wormhole through the island region\cite{penington2019replica,almheiri2020replica}. 

For an evaporating black hole coupled with a flat space bath, it was shown that for a large enough region in the radiation ({\it i.e.} the bath), the entanglement island becomes nontrivial after a time known as the Page time. Taking into account of the entanglement island, the entanglement entropy of the radiation becomes smaller than the thermal value in Hawking's calculation, which correctly resolves the conflict between gravitational calculation and unitarity. Following the entanglement wedge reconstruction in holographic duality, the island formula also suggests that (small perturbations in) the entanglement island region of the black hole interior can be reconstructed in the radiation. Retrieval of information from the entanglement island has been discussed using different methods\cite{penington2019replica,chen2020pulling}.

In this paper, we want to address the following questions: In general, how difficult is the information retrieval from the entanglement island? Is it possible to carry a measurement in the radiation that measures the state of an interior qubit in the entanglement island? How will measurements in the radiation affects the physics seen by an infalling observer? We begin by proving a general result on state reconstruction. Similar to the bulk-boundary correspondence of relative entropy shown in Ref. \cite{jafferis2016relative}, we use replica trick to show that the ``boundary'' one-norm distance between two (similar) states in a bath region $A$ is equal to the ``bulk" one-norm distance between corresponding states in the region $AI$, if $A$ has an entanglement island $I$. This is a straightforward generalization of the AdS/CFT results where boundary region is replaced by a region in the bath, and the entanglement wedge is replaced by the entanglement island. As a consequence of this general state reconstruction formula, we show that a region in the bath that is only classically correlated with the rest of the system can never reconstruct information that is encoded to the system by applying a unitary in the interior, because the interior is space-like separated from the bath. However, this leads to a contradiction. If there is a region $A$ with nontrivial entanglement island $I$, and another region $W$ without entanglement island, we can carry a measurement on $A$ that retrieves the information in the interior region $I$, and send the measurement result to $W$. On the other hand, after the measurement (which is a unitary acting on the bath system), we show that the state reconstruction formula suggests the state of $W$ still has no correlation with the interior information, which is contradictory with the fact that $W$ knows the measurement result from $A$. By contradiction this suggests that the replica wormhole calculation that leads to the state reconstruction formula cannot apply to both regions discussed above. We propose that whenever an interior-information-retrieval measurement occurs in the bath, the state reconstruction formula fails for the state after the measurement. (Note that we always understand "measurement" as a unitary operator applying to $A,W$ and other parts of the bath, so the problem is not caused by non-unitarity.) 

We name such measurement operators as ``miracle operators'' since they have to change the gravitational path integral in a nontrivial way. 
By explicitly constructing miracle operators using a replica trick, we understand why the original state reconstruction formula fails when such an operator is inserted. Additional replica wormholes arise not between different replicas of the original black hole, but between the original black hole and ``simulated black hole''\footnote{We thank Ahmed Almheiri for suggesting this term.} used for defining the miracle operator. Using this explicit construction, we obtain some new understanding to the firewall paradox\cite{almheiri2013black,almheiri2013apologia} in two different setups. We show that a measurement in the radiation can measure the state of an interior qubit, and an infalling obsever will see that the entanglement of two qubits across the quantum extremal surface can be destroyed by such exterior measurements. Furthermore, we construct a different measurement which applies to the radiation and checks that one of the radiation qubit is in a particular entangled state with an interior qubit in the entanglement island. In this setup, an infalling observer will still be able to verify the entanglement of the entangled pair, which appears to contradict with the expectation in ordinary quantum mechanics. This is allowed because the entanglement checking operator must come with its own gravitational theory and provide additional copies of the universe including the infalling observer. Interestingly, in the replica calculation of this setup, one can see that the success of entanglement checking is related to the nontrivial homotopy group of the replica manifold.

The remainder of the paper is organized as follows. In Sec. \ref{sec:reconstruction} we derive a state reconstruction formula for the evaporating black hole system based on the results of entanglement island and replica wormhole. In Sec. \ref{sec:paradox} we show that there is an apparent paradox caused by applying the quantum extremal surface formula to general dynamics of the radiation. In Sec. \ref{sec:miracle} we discuss the how the paradox is resolved by realizing that some operators can introduce extra copies of geometry and wormholes connecting with them, which we name as miracle operators. Sec. \ref{sec:firewall} applies the miracle operator to address the firewall paradox in two different setups. Finally Sec. \ref{sec:conclusion} contains a summary and some further discussion.

\section{State reconstruction formula for an evaporating black hole}
\label{sec:reconstruction}

We begin by an overview of the island formula, from the point of view of replica calculation.\cite{almheiri2020replica,penington2019replica} For concreteness we consider a single-sided AdS black hole formed from collapse, which is coupled with a flat space bath. Taking a subsystem $A$ of the bath, the computation of ${\rm tr}\left(\rho_A^n\right)$ involves taking $n$ replicas of the entire system, and introducing a cyclic permutation of the different replicas in region $A$. The geometry of the system is a union $\mathcal{B}_g\cup \mathcal{R}_n$, with $\mathcal{R}_n$ the flat space with branch covering at the boundary of $A$, and $\mathcal{B}_g$ the geometry of the AdS black hole. The gravitational path integral is defined for the metric of $\mathcal{B}_g$, with the boundary condition fixed by $\mathcal{R}_n$. 
\begin{align}
    Z_n(A)=\int_{\mathcal{B}_g}Dg\int_{\mathcal{B}_g\cup\mathcal{R}_n}D\phi e^{-\mathcal{A}_{EH}[g]-\mathcal{A}_{QFT}[\phi,g]}
\end{align}
Here $\phi$ represents all matter fields. (More precisely, we should introduce $n+m$ replicas and introduce the twist only in the first $n$ copies, and in the end take $m\rightarrow-n$, as is discussed in Ref. \cite{dong2020effective}. The $m$ replicas can be omitted if we assume no saddle point will contain wormhole connecting them with each other or with the first $n$ replicas.) Assuming the path integral is dominated by a classical saddle point which preserves the replica symmetry $Z_n$, the QFT contribution is given by the $n$-th Renyi entropy of the matter fields in a region $IA$, with $I$ in the gravitational bulk. The gravitational contribution $\mathcal{A}_{EH}[g_{saddle}]$ can be computed by taking a quotient $\mathcal{B}_g/Z_n$ and evaluate the Einstein-Hilbert action. If there is a nontrivial replica wormhole in region $I$, the boundary of $I$ becomes a conical singularity with co-dimension $2$ and conical angle $\frac{2\pi}n$ in the quotient manifold $\mathcal{B}_g/Z_n$. In the limit $n\rightarrow 1$, this leads to the quantum extremal surface formula
\begin{align}
    -\log Z_n(A)&\simeq \mathcal{A}_{EH}[g_{saddle}]+(n-1)S^{(n)}_{QFT}(IA)\\
    S(A)&\simeq {\rm ext}_I\left[\frac{\left|\partial I\right|}{4G_N}+S_{QFT}(IA)\right]\label{eq:QES}
\end{align}
with ${\rm ext}_I$ referring to extremization over choice of spatial region $I$. If there are multiple extremal surfaces, the one with minimal $S(A)$ should be chosen. 

\begin{figure}
    \centering
    \includegraphics[width=4in]{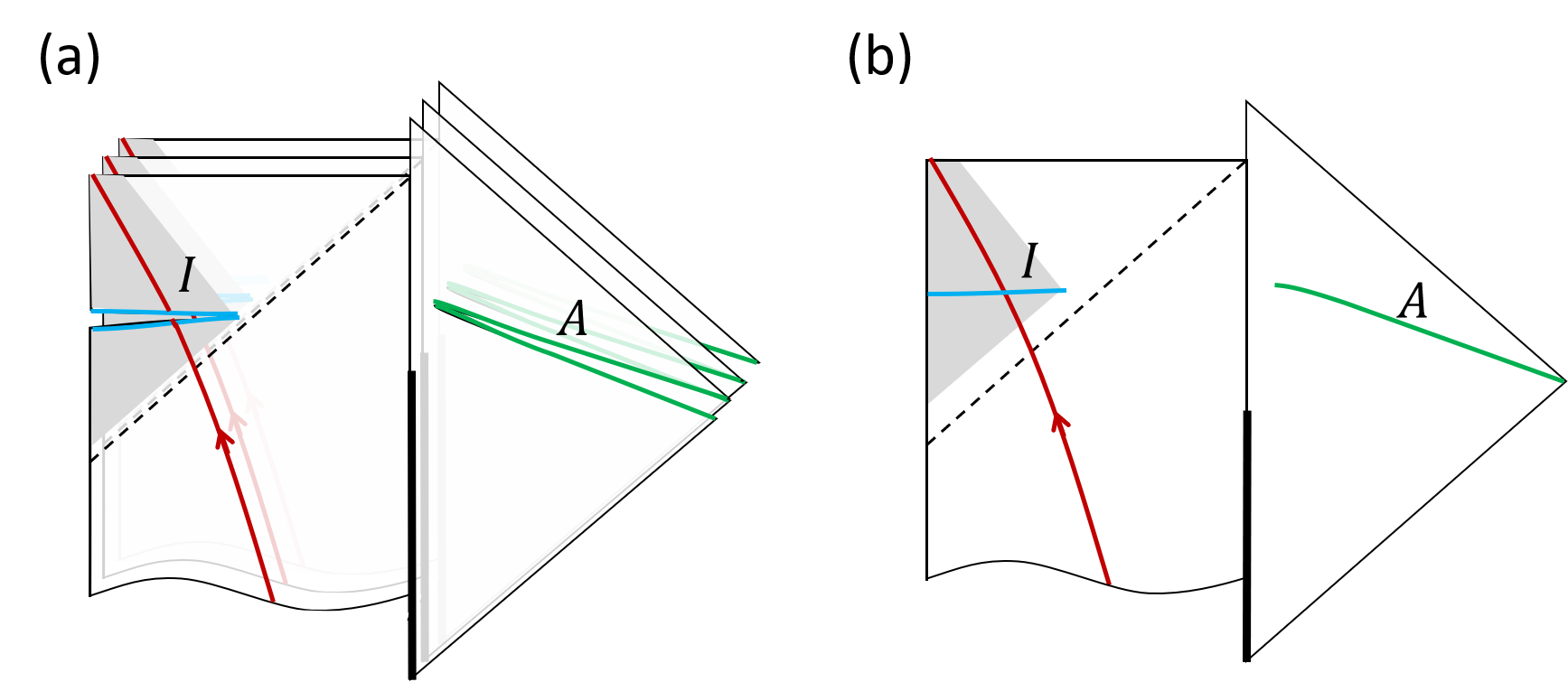}
    \caption{(a) Illustration of the replica manifold with branch-covering at $A$ and possibly an island region $I$. (b) In the analytic continuation to $n\rightarrow 1$ limit, illustration of the quantum extremal surface (\ref{eq:QES}) which is the boundary of $I$. }
    \label{fig:island}
\end{figure}

The discussion here is entirely parallel with the proof of the HRT formula in ordinary AdS/CFT\cite{faulkner2013quantum,lewkowycz2013generalized,dong2016deriving,dong2018entropy}, except that the gravitational theory now has a different boundary condition set by the radiation. We would like to emphasize that the discussion seems independent from the detail of dynamics of the radiation. Whether the radiation is a flat space CFT or a quantum computer, we expect the quantum extremal surface derivation above to hold, as long as the gravity in the AdS region remains semi-classical.

In addition to the entropy formula, the replica calculation can also tell us more about how operator reconstruction works. We first clarify some notation. Denote the Hilbert space of the holographic CFT as $\mathbb{H}_B$, and that of the bath as $\mathbb{H}_R$ ($R$ for ``radiation''), the Hilbert space of the entire system is $\mathbb{H}_{QG}=\mathbb{H}_B\otimes\mathbb{H}_R$.\footnote{Of course in a more realist system with dynamical gravity everywhere, the Hilbert space does not factorize. The states $\rho_A,\sigma_A$ defined below should be viewed as ``effective" quantum states\cite{dong2020effective} describing correlations of quantum field theory degrees of freedom of the radiation qubits. It is unclear how to rigorously define this if gravity is dynamical in the bath. } We will refer to this Hilbert space as the quantum gravity Hilbert space. A classical saddle point is denoted by $\mathcal{M}$, which is a manifold that includes the asympotic AdS part and the attached flat space part. For each $\mathcal{M}$ (with Lorenzian signature) and a given Cauchy surface, the path integral defines a state of the quantum field theory on that background. We denote the Hilbert space of the quantum field theory as $\mathbb{H}_{QFT}$. $\mathbb{H}_{QG}$ and $\mathbb{H}_{QFT}$ are the analog of boundary Hilbert space and bulk Hilbert space in AdS/CFT without bath. In the rest of the paper, we will use bold font Greek letters $\rhoqft,\sigmaqft$ to denote states in $\mathbb{H}_{QFT}$, and use regular font Greek letters $\rho,\sigma$ to denote those in the boundary-and-bath Hilbert space $\mathbb{H}_{QG}$.

Now we consider two states $\rhoqft,\sigmaqft\in\mathbb{H}_{QFT}$. When these two states are close to each other, such that the energy momentum tensor difference is order $O(G_N^0)$, the back reaction to geometry is small. In the replica calculation, we can compute quantity such as ${\rm tr}\left(\rho^{n-m}_A\sigma^m_A\right)$ and expect that it is dominated by the same saddle point as ${\rm tr}\left(\rho^n_A\right)$. In this case one obtains
\begin{align}
    {\rm tr}\left(\rho_A^{n-m}\sigma_A^m\right)\simeq e^{-\mathcal{A}_n}{\rm \bf tr}\left(\rhoqft^{n-m}_{AI}\sigmaqft^m_{AI}\right)
;\end{align}
with $\mathcal{A}_n$ the contribution of the Einstein-Hilbert action at the saddle point manifold. Consequently, we obtain
\begin{align}
    \left\lVert \rho_A-\sigma_A\right\rVert_{2n}&= {\rm tr}\left[\left(\rho_A-\sigma_A\right)^{2n}\right]\simeq e^{-\mathcal{A}_{2n}}\left\lVert\rhoqft_{AI}-\sigmaqft_{AI}\right\rVert_{2n}\label{eq:2n norm}
\end{align}
for positive integer $n$. In the analytic continuation to $n\rightarrow 1/2$, $\mathcal{A}_{2n}$ vanishes:
\begin{align}
    \mathcal{A}_{2n}\simeq (2n-1)\frac{\left|\partial I\right|}{4G_N}
\end{align}
Therefore we obtain
\begin{align}
    \left\lVert \rho_A-\sigma_A\right\rVert_{1}=\left\lVert\rhoqft_{AI}-\sigmaqft_{AI}\right\rVert_1\nonumber 
\end{align}
This discussion is a simple generalization of Ref. \cite{jafferis2016relative} for relative entropy (see also \cite{almheiri2020replica,chen2020pulling} for related discussions). Alternatively, we could also prove Eq. (\ref{eq:one norm}) using the conclusion of Ref. \cite{jafferis2016relative}: Since the mapping $\mathcal{C}: \rhoqft_{AI}\rightarrow \rho_A$ preserves the relative entropy between two states $\rho,\sigma$ in the code subspace, it is invertable and thus preserves the trace distance between two states. 

\begin{figure}
    \centering
    \includegraphics[width=4in]{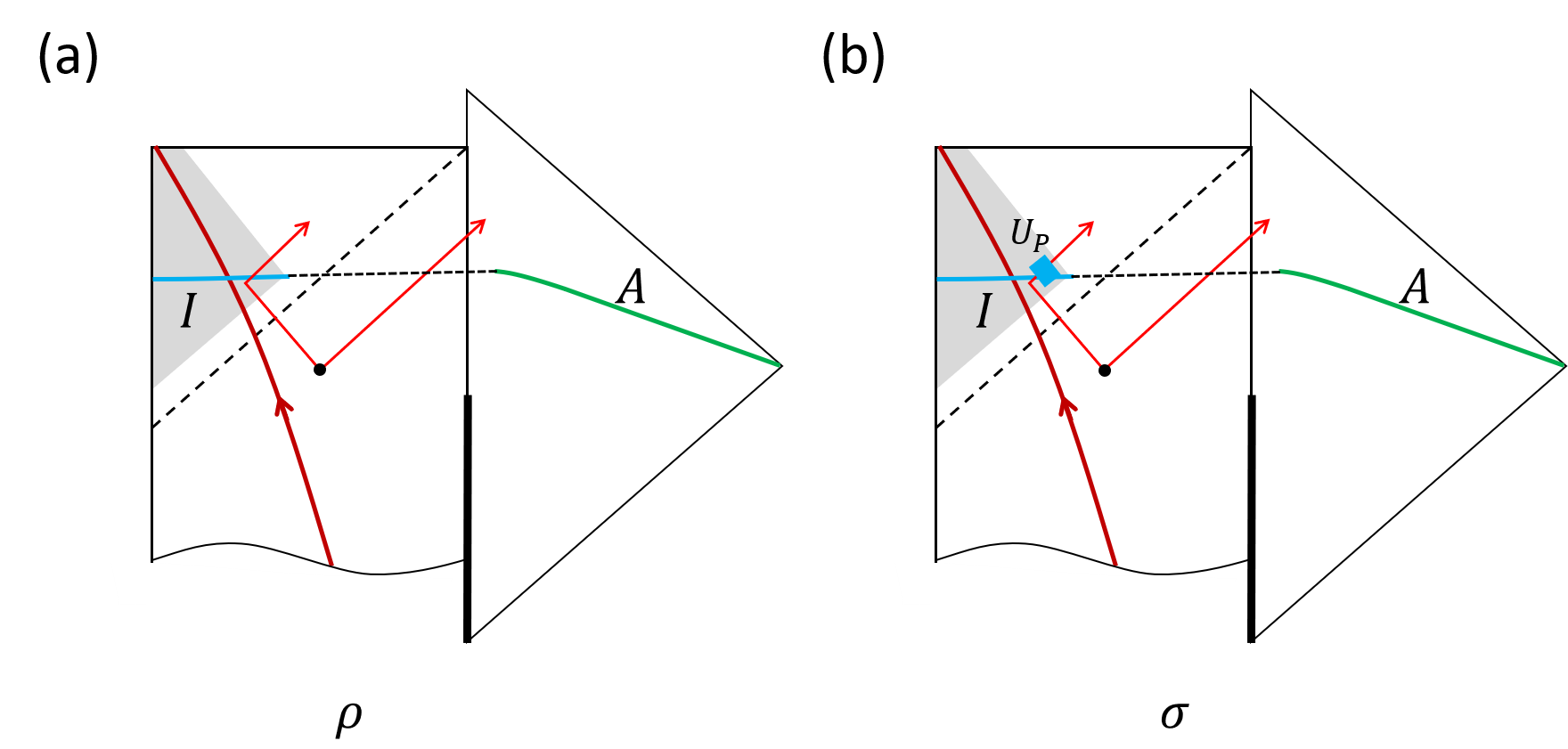}
    \caption{Illustration of two states $\rho,\sigma$ which are only different by a local unitary $U_P$ acting on a small region $P$ in the island. For example $U_P$ can flip the spin of a particle. The QFT states are defined on a Cauchy surface that includes $A$ and $I$.}
    \label{fig:rhosigma}
\end{figure}
This result plays a central role in this paper, so we would like to summarize it in the following theorem:
\begin{theorem}\label{thm:theorem1}
{\bf State reconstruction formula.} For two states $\rho,\sigma\in\mathbb{H}_{QG}$, assume that the gravitational path integral in the calculation of ${\rm tr}\left(\rho_A^n\right)$ and ${\rm tr}\left(\sigma_A^n\right)$ are both dominated by the same smooth manifold $\mathcal{M}_n$ preserving the cyclic permutation symmetry between replicas, up to corrections $O(G_N)$, then 
\begin{align}
    \left\lVert \rho_A-\sigma_A\right\rVert_{1}=\left\lVert\rhoqft_{AI}-\sigmaqft_{AI}\right\rVert_1\label{eq:one norm}
\end{align}
with $I$ the entanglement island of $A$.
\end{theorem}

As an example, consider $\rhoqft$ as the state of an evaporating black hole coupled with the bath, in which a region of the radiation $A$ has an entanglement island $I$. $\sigmaqft=U_P\rhoqft U_I^\dagger$ is different from $\rhoqft$ by a local unitary acting in a small spacetime region $P\in I$, as is illustrated in Fig. \ref{fig:rhosigma}. Eq. (\ref{eq:one norm}) tells us that the difference between $\rho$ and $\sigma$ is preserved in the radiation region $A$. For example, if $U_P$ rotates the spin of a qubit from $z$-direction eigenstates up to down, one can measure the spin z-component operator $Z_P$ of this qubit to distinguish $\rhoqft_{AI}$ and $\sigmaqft_{AI}$. Then Eq. (\ref{eq:one norm}) guarantees that there is a projection operator $P_A$ in $A$ such that
\begin{align}
    \left|{\rm tr}\left(P_A\rho_A\right)-{\rm tr}\left(P_A\sigma_A\right)\right|&=\frac12\left\lVert \rho_A-\sigma_A\right\rVert_{1}\nonumber\\
    &=\frac12\left\lVert\rhoqft_{AI}-\sigmaqft_{AI}\right\rVert_1\nonumber\\
    &\geq \frac12\left|{\rm tr}\left(Z_P\rhoqft_{AI}\right)-{\rm tr}\left(Z_P\sigmaqft_{AI}\right)\right|\label{eq:operator}
\end{align}
Thus Eq. (\ref{eq:one norm}) guarantees that $\rho_A,\sigma_A$ can be distinguished by a measurement in region $A$, to the same degree that $\rhoqft_{AI}$ and $\sigmaqft_{AI}$ can be distinguished on $AI$. In the next section, we will discuss a more explicit setup for extracting such information, which leads to an apparent contradiction.

\section{A paradox in state reconstruction}\label{sec:paradox}

\subsection{A no-go theorem}

\begin{figure}
    \centering
    \includegraphics[width=4in]{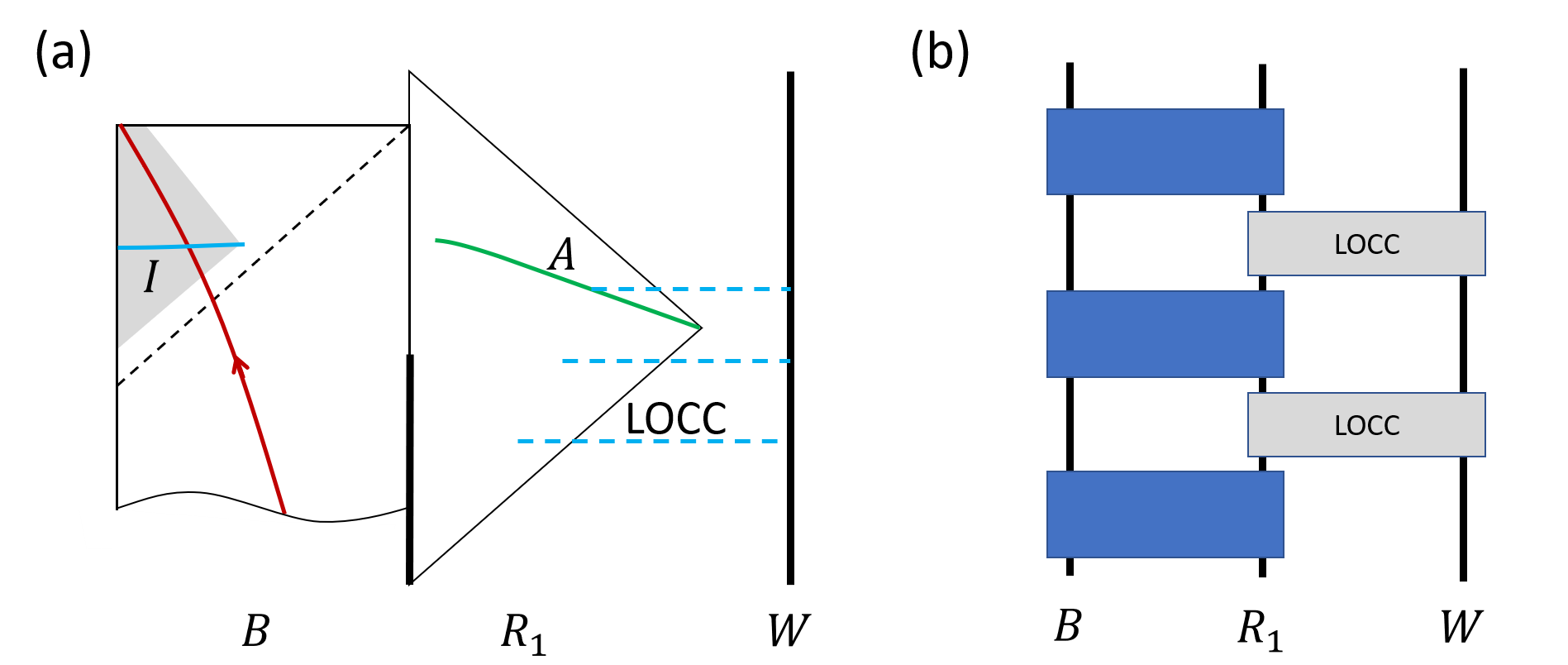}
    \caption{(a) Illustration of the ancilla $W$ which only couples with the rest of the bath $R_1$ through LOCC. (b) A quantum circuit representation of the same setup, with black hole $B$ couples with $R_1$ through quantum gates, while $R_1$ and $W$ are coupled only by LOCC.}
    \label{fig:locc}
\end{figure}

As we discussed earlier, the island formula (\ref{eq:QES}) only requires the geometry of the gravitational system to have small fluctuation, independent from the dynamics of the radiation. In this subsection, we will consider a special class of radiation system, as is shown in Fig. \ref{fig:locc}. The radiation system $R$ consists of two subsystems $R_1$ and $W$. $R_1$ is the ``ordinary" radiation system, which is a flat space QFT coupled with the black hole. $W$ is an ancilla that only couples with $R_1$ by local operations and classical communication (LOCC). The initial state of 
$R$ is a direct product state $\rho_{0R_1}\otimes \rho_{0W}$.  Physically, we can think $W$ as a model of lab equipment of an observer we use to measure the radiation. We assume the observer only has access to the radiation system $R_1$ through LOCC, which physically involves multiple rounds of carrying quantum measurements to $R_1$ and applying unitaries on it. We include the definition of LOCC in Appendix \ref{app:LOCC}. This setup is an example of an incoherent quantum algorithmic measurement (QUALM) defined in Ref. \cite{aharonov2021quantum}.  

Now we study the entropy of the $W$ subsystem. With the assumption that the island formula applies, we obtain
\begin{align}
    S(W)={\rm ext}_I\left[\frac{\left|\partial I\right|}{4G_N}+S_{QFT}(WI)\right]
\end{align}
Empty $I=\emptyset$ is always a saddle point, with contribution $S_{QFT}(W)$. If $W$ has a nontrivial entanglement island $I$, the necessary condition is 
\begin{align}
    S_{QFT}(WI)<\frac{\left|\partial I\right|}{4G_N}+S_{QFT}(WI)<S_{QFT}(W)
\end{align}
However, in the situation we consider, $W$ and the remainder of the system ($R$ and black hole $B$) is coupled with LOCC, such that the state remains separable at all time:
\begin{align}
    \rho_{QFT}=\sum_ip_i\rho_{QFT}^i(BR)\rho_{QFT}^i(W)
\end{align}
Therefore for any subsystem $I$, $\rho_{QFT}(IW)$ is still separable. Consequently, the conditional entropy is always non-negative (see Appendix \ref{app:conditional entropy}):
\begin{align}
    H_{QFT}(I|W)\equiv S_{QFT}(WI)-S_{QFT}(W)\geq 0
\end{align}
Consequently, $W$ always has a trivial island. We summarize this conclusion in the following theorem:

\begin{theorem}\label{thm:main}
If i) a subsystem $W$ of the radiation only couples with the remainder of the system (including the radiation and the black hole) through LOCC, and the initial state is separable between $W$ and its complement, and ii) semiclassical approximation applies to the entropy calculation of $W$, then $W$ has no entanglement island.
\end{theorem}

\subsection{An apparent paradox}\label{subsec:paradox}

Theorem \ref{thm:theorem1} and Theorem \ref{thm:main} both seem to be general and are natural consequences of the island formula, but in the following we will show that there is a contradiction when we try to apply them to a particular setup.


\begin{figure}
    \centering
    \includegraphics[width=4in]{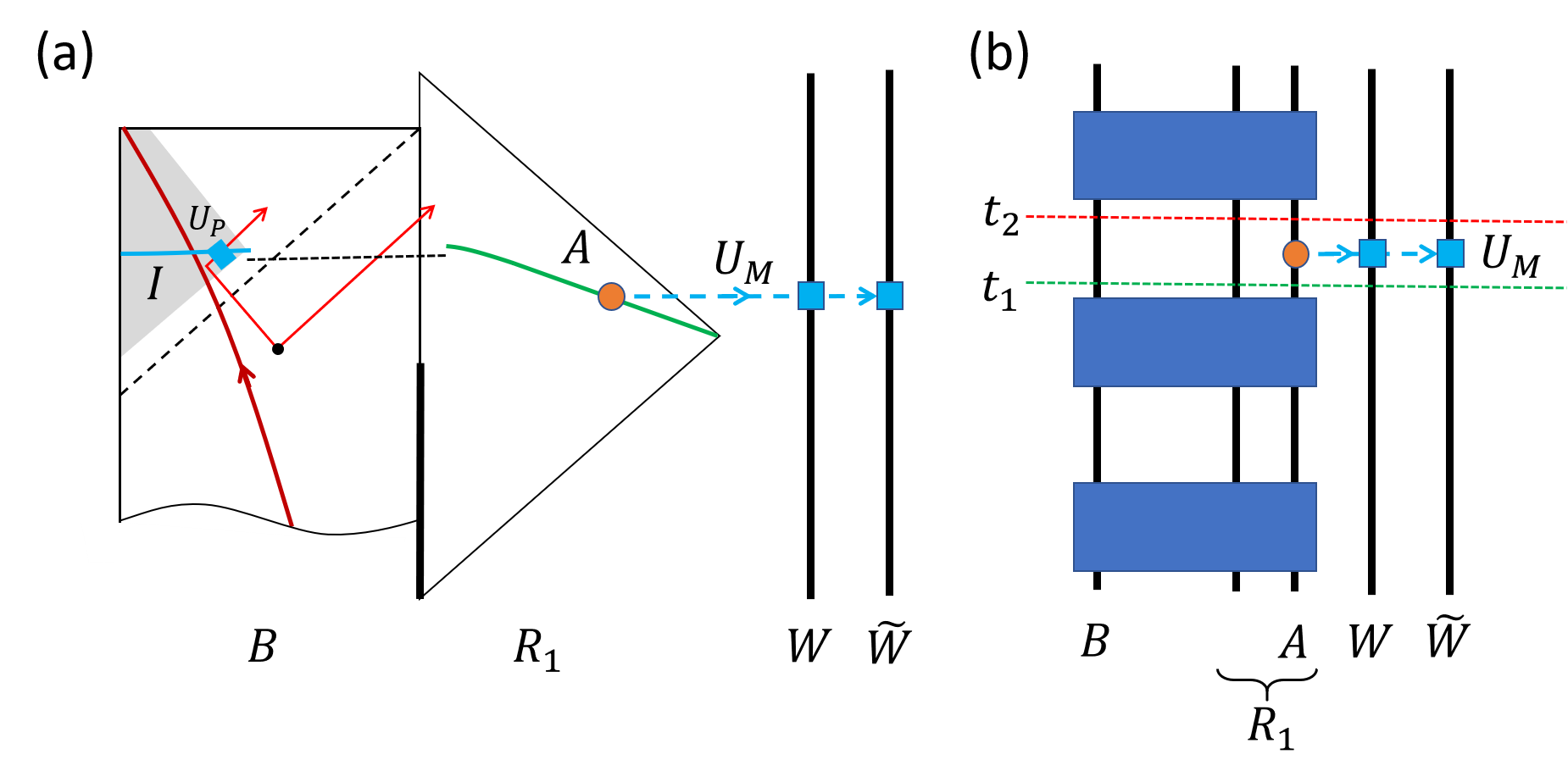}
    \caption{Illustration of the particular setup in subsection (\ref{subsec:paradox}) in Penrose diagram (a) and quantum circuit (b). We consider two states different by an interior unitary $U_P$, as was discussed in Fig. \ref{fig:rhosigma}. In addition, a unitary $U_M$ (defined by Eq. (\ref{eq:measurement unitary}) applied to region $A$ and two ancilla qubits $W,\tilde{W}$ measures $A$ and records the result on $W$. The green and red horizontal dashed lines indicate the time $t_1$ before applying $U_M$, and $t_2$ right after applying $U_M$. What is relevant to our discussion is the application of state reconstruction formula (\ref{eq:one norm}) for $A$ at time $t_1$ and $W$ at time $t_2$. }
    \label{fig:measurement}
\end{figure}

Consider the setup we discussed earlier in Fig. \ref{fig:rhosigma}, with two states $\rho$ and $\sigma=U_P\rho U_P^\dagger$ that are only different by a unitary applying to a small region in the island $I$ of a radiation region $A$. In addition, we consider an ancilla $W$ that is coupled with $A$ through the following quantum channel:
\begin{align}
    \mathcal{C}_M\left(\rho_A\otimes \ket{0_W}\bra{0_W}\right)=P_A\rho_AP_A\otimes \ket{1_W}\bra{1_W}+\left(\mathbb{I}_A-P_A\right)\rho_A\left(\mathbb{I}_A-P_A\right)\otimes \ket{0_W}\bra{0_W}\label{eq:measurement channel}
\end{align}
Here $P_A$ is the particular projection operator that distinguish $\rho_A$ and $\sigma_A$ optimally, such that ${\rm tr}\left(P_A\rho_A\right)-{\rm tr}\left(P_A\sigma_A\right)=\frac12\left\lVert \rho_A-\sigma_A\right\rVert_{1}$. The channel $\mathcal{C}_M$ measures $A$ with the projector $P_A$, and save a copy of the measurement result on $W$. $\mathcal{C}_M$ is an LOCC. We would like to emphasize that although $\mathcal{C}_M$ is not a unitary operator, it could be realized by an unitary in a slightly bigger system. For example if we introduce another qubit $\tilde{W}$, we can define the unitary on $AW\tilde{W}$:
\begin{align}
    U_M=P_A\otimes X_W\otimes X_{\tilde{W}}+\left(\mathbb{I}_A-P_A\right)\otimes \mathbb{I}_W\otimes\mathbb{I}_{\tilde{W}}\label{eq:measurement unitary}
\end{align}
with $X_W$ and $X_{\tilde{W}}$ the Pauli $x$ operator that flips between $\ket{0}$ and $\ket{1}$ states. Applying $U_M$ to the state $\rho_A\otimes\ket{0_W}\bra{0_W}\otimes\ket{0_{\tilde{W}}}\bra{0_{\tilde{W}}}$ and tracing over $\tilde{W}$ leads to $\mathcal{C}_M\left(\rho_A\otimes\ket{0_W}\bra{0_W}\right)$. We would like to emphasize this point to clarify that introducing $\mathcal{C}_M$ in the discussion does not imply we have violated unitarity in $R$, as long as $A,W,\tilde{W}$ are all part of the bath system $R$. The setup is illustrated in Fig. \ref{fig:measurement}. 

Now we apply the state reconstruction formula (Theorem \ref{thm:theorem1}) to two different regions.
\begin{enumerate}
    \item For the state before $U_M$ is applied, defined at time $t_1$ in Fig. \ref{fig:measurement}, applying the state reconstruction formula to region $A$ leads to $\norm{\rho_A-\sigma_A}_1=\norm{\rhoqft_{AI}-\sigmaqft_{AI}}_1$, which is order $1$.
    \item For the state after $U_M$ is applied, defined at time $t_2$ in Fig. \ref{fig:measurement}, applying the state reconstruction formula to $W$ leads to 
    \begin{align}
        \norm{\rho_W-\sigma_W}_1=\norm{\rhoqft_W-\sigmaqft_W}_1\label{eq:zero norm},
    \end{align}
    since $W$ still has only classical correlation with the rest of the system, and thus has no entanglement island according to Theorem \ref{thm:main}. Furthermore, remember that $\rhoqft_W$ and $\sigmaqft_W$ are the states of $W$ in QFT, which are obtained by fixing the geometry and defining the QFT state on a Cauchy surface by a QFT path integral. Therefore a unitary $U_P$ that is acting in a spacetime region spacelike separated from $W$ will have trivial effect on the reduced state of $W$, leading to $\rhoqft_W=\sigmaqft_W$.  Therefore Eq. (\ref{eq:zero norm}) implies $\left\lVert \rho_W-\sigma_W\right\rVert_1=0$.
\end{enumerate}

However, the two conclusions above are in direct contradiction, because $W$ has learned about the measurement result from $A$, and therefore ``inherited'' a nontrivial one-norm distance from that between $\rho_A$ and $\sigma_A$. To see that, we write down the state of $W$ after the coupling:
\begin{align}
    \rho_W&={\rm tr}_A\left[\mathcal{C}_M\left(\rho_A\otimes \ket{0_W}\bra{0_W}\right)\right]=p_\rho \ket{1_W}\bra{1_W}+(1-p_\rho)\ket{0_W}\bra{0_W}\\
    \text{with~}p_\rho&={\rm tr}\left(P_A\rho_A\right)
\end{align}
Similarly, when the state of $A$ before measurement is $\sigma_A$, the state of $W$ after measurement is $\sigma_W=p_\sigma \ket{1_W}\bra{1_W}+(1-p_\sigma)\ket{0_W}\bra{0_W}$ of the same form, with $p_\sigma={\rm tr}(P_A\sigma_A)$. Therefore we have
\begin{align}
    \left\lVert \rho_W-\sigma_W\right\rVert_1=2\left|p_\rho-p_\sigma\right|=\left\lVert \rho_A-\sigma_A\right\rVert_1\label{eq:equal norm}
\end{align}
Physically, Eq. (\ref{eq:equal norm}) tells us that one-norm distance is transferable by classical communication, because it is a measure of {\it classical} information one can learn by an optimal measurement. This transferability thus suggests $\left\lVert \rho_W-\sigma_W\right\rVert_1$ should be order $1$, in direct conflict with the result of state reconstruction formula applied to $W$. Therefore we conclude that to avoid inconsistency, at least for in one of the two states we discussed (the state of $A$ at $t_1$ and the state of $W$ at $t_2$), the assumptions of Theorem \ref{thm:theorem1} must fail.

We will discuss more about the interpretation of this apparent paradox in next section.

\section{The miracle operators}
\label{sec:miracle}

Let us summarize the problem again. The replica calculation that leads to the QES formula also predicts Eq. (\ref{eq:one norm}), which tells us that small perturbation in $I$ can be reconstructed in $A$. On the other hand, the same formula suggests that an ancilla which only couples to the black hole and radiation by LOCC cannot probe such small perturbation anywhere space-like separated from $A$ and the ancilla, because it does not have an entanglement island (Theorem \ref{thm:main}). The only assumption we have used to achieve this paradox is the assumptions of Theorem \ref{thm:theorem1}, that the gravitational path integral is dominated by a smooth saddle point manifold in the replica calculation of both $S_A$ and $S_W$. (Note that we do not need to assume the saddle point manifold to be the same one for these two calculations. The paradox remains as long as the manifolds involved are smooth and replica symmetric, even if we are allowed to consider a large back-reaction caused by $U_M$.) Therefore by contradiction we have proved that this assumption must fail for at least one of the calculation. If we assume the $S_A$ calculation is correct (which occurs before $W$ got involved), then we have to conclude that {\it after applying $U_M$, the calculation of ${\rm tr}\left(\rho_W^n\right)$ is not dominated by any smooth replica symmetric saddle point manifold.}  The problem remains if we replace a single saddle point with a sum over multiple saddle points. Since the contribution of each possible saddle point (for the calculation of $W$ Renyi entropy) with a nontrivial island $I$ is suppressed by a factor $e^{-\Delta S}$ with $\Delta S\sim O\left(1/G_N\right)$, the contribution remains suppressed even if we sum over polynomial (in $1/G_N$) number of saddle points. In other words, applying the particular unitary $U_M$ to the bath system and computing ${\rm tr}\left(\rho_W^n\right)$ (which can be interpreted as measuring the expectation value of a twist operator in the $n$-replica system) change the behavior of the gravitational path integral in a way that is much more dramatic than an ordinary back-reaction. From our discussion one can see that this problem occurs whenever we are carrying a measurement such as $P_A$, which reveals nontrivial information about the interior. In the following we will refer to such interior-information-revealing measurement operators as ``miracle operators'', due to their dramatic effect to the spacetime geometry.

\begin{figure}
    \centering
    \includegraphics[width=5.2in]{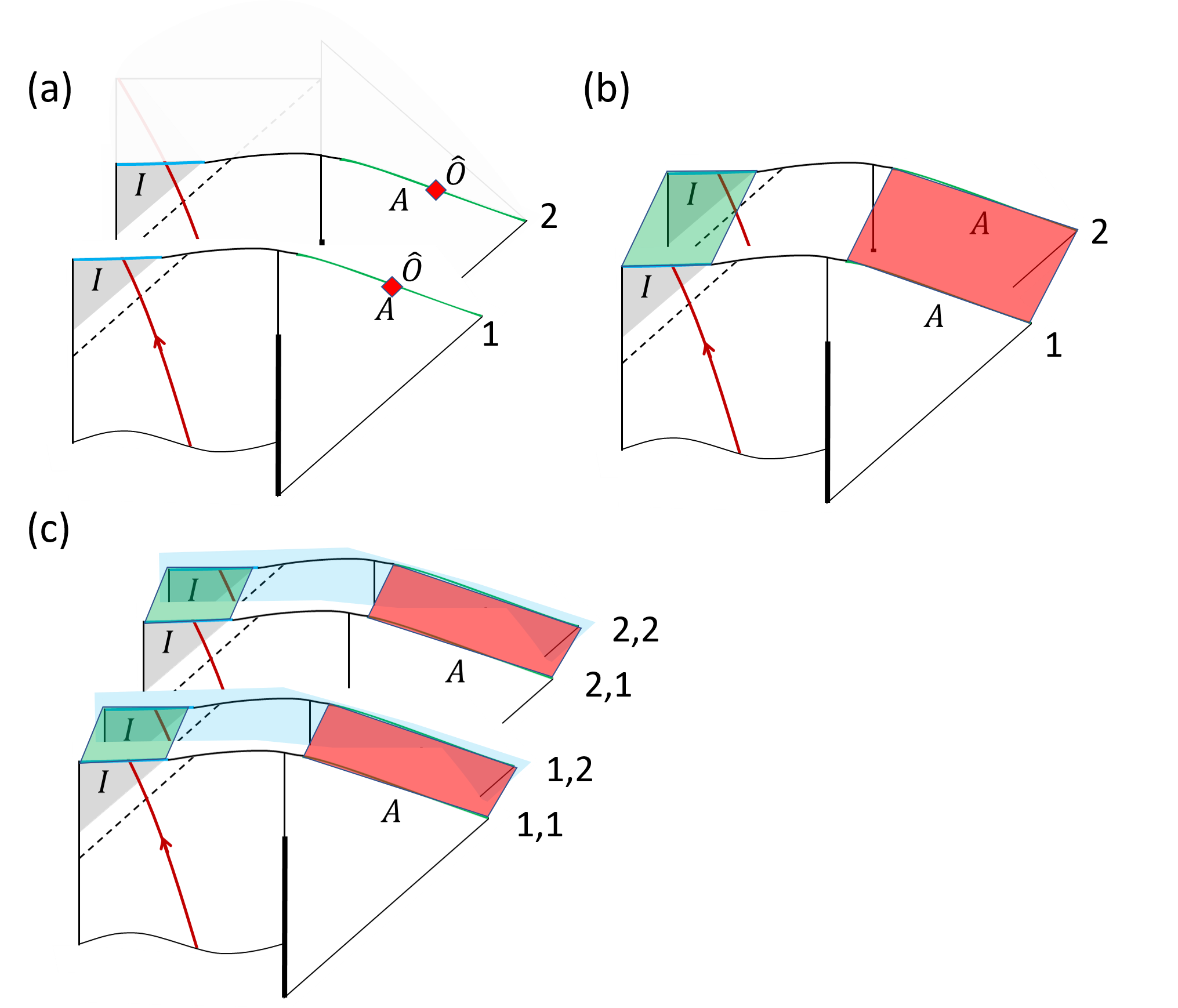}
    \caption{Illustration of the replica calculation for (a) ${\rm tr}(\rho_A\hat{O})^n$ for a regular operator $\hat{O}$, (b) ${\rm tr}\left(\rho_A^n\right)$ and (c) ${\rm tr}\left(\rho_AZ_A^{(k)}\right)^n$ with $Z_A^{(k)}$ defined in Eq. (\ref{eq:def Zk}). The illustration shows the case $n=2,k=1$. The part of Penrose diagram below a Cauchy surface represents the QFT state at that surface, prepared by the QFT path integral on the given background geometry. The red bridge connecting different replicas in (b) and (c) is introduced by multiplying $\rho_A$ with $\rho_A$ or $Z_A^{(k)}$. The green bridge is the replica wormhole. Note that in (c) replica wormhole connects the physical copies $(1,1)$ and $(2,1)$ with simulated copies $(1,2)$ and $(2,2)$ in blue, which represent $Z_A^{(k)}$.}
    \label{fig:replica Zk}
\end{figure}

To understand what happens when we apply miracle operators, we write down explicitly the form of the optimal measurement $P_A$. For any two states $\rho_A,\sigma_A$, if we write $\rho_A-\sigma_A$ in the diagonal basis $\rho_A-\sigma_A=\sum_n\lambda_n\ket{n}\bra{n}$, then
\begin{align}
    P_A=\sum_n\frac12\left(1+{\rm sgn}\left(\lambda_n\right)\right)\ket{n}\bra{n}\label{eq:PA definition}
\end{align} 
with ${\rm sgn}(\lambda_n)=+1,0,-1$ for $\lambda_n>0,\lambda_n=0,\lambda_n<0$, respectively.\footnote{If $\rho_A-\sigma_A$ is not full rank, $P_A$ has eigenvalues $0,1$ and $\frac12$, so that it is not a projector. Correspondingly there are three possible measurement results $W$ needs to record. However, when the input state is $\rho_A$ or $\sigma_A$, only two measurement results will appear. All our discussion remains valid if we allow a general input state and define $W$ to have Hilbert space dimension $3$.} Therefore we can introduce the replica trick and express $P_A$ as:
\begin{align}
    2P_A-\mathbb{I}_A&={\rm sgn}\left(\rho_A-\sigma_A\right)=\left.\left(\rho_A-\sigma_A\right)^{2n-1}\right|_{n\rightarrow \frac12}\label{eq:operator replica}
\end{align}
from which we can see explicitly that $P_A$ involves gravitational path integral. Consequently, the evaluation of $P_A$ can involve replica wormholes which connect the interior of the original black hole and those in $P_A$. The operators reconstructed using Petz map\cite{penington2019replica} are also examples of miracle operators. The operator we consider is simpler than the Petz map case because we focus on a simpler task of distinguishing two particular states.

Using this replica trick, we can see the reason of the contradiction we find in last section. For this purpose, define 
\begin{align}
    Z_A^{(k)}=\left(\rho_A-\sigma_A\right)^{2k-1}\label{eq:def Zk}
\end{align}
for integer $k$, and introduce a $W$ that is coupled to $A$ by the following LOCC channel:
\begin{align}
    \mathcal{C}_M^{(k)}\left(\rho_A\otimes\ket{0_W}\bra{0_W}\right)&=X^{(k)}_{A+}\rho_AX^{(k)}_{A+}\otimes\ket{1_W}\bra{1_W}+X^{(k)}_{A-}\rho_AX^{(k)}_{A-}\otimes\ket{0_W}\bra{0_W}\label{eq:replica channel}\\
    \text{with}~X_{A\pm}^{(k)}&=\sqrt{\frac12\left(\mathbb{I}_A\pm Z_A^{(k)}\right)}
\end{align}
Note that $\mathbb{I}_A\pm Z_A^{(k)}$ is positive, so that the square root operator $X_{A\pm}^{(k)}$ is uniquely defined and Hermitian.
This channel carries a positive operator-valued measurement (POVM) to $A$ and store the result in $W$. After applying the channel, the state of $W$ is
\begin{align}
    \rho_W&=p_\rho^{(k)}\ket{1_W}\bra{1_W}+\left(1-p_\rho^{(k)}\right)\ket{0_W}\bra{0_W}\\
    p_\rho^{(k)}&=\frac12\left(1+{\rm tr}\left(Z_A^{(k)}\rho_A\right)\right)
\end{align}
Now if we compute ${\rm tr}\left(\rho_W^n\right)$, we obtain
\begin{align}
   {\rm tr}\left(\rho_W^n\right)&=2^{-n}\left[\left(1+{\rm tr}\left(Z_A^{(k)}\rho_A\right)\right)^n+\left(1-{\rm tr}\left(Z_A^{(k)}\rho_A\right)\right)^n\right]
\end{align}
which involves ${\rm tr}\left(Z_A^{(k)}\rho_A\right)^m={\rm tr}\left(\left(\rho_A-\sigma_A\right)^{2k-1}\rho_A\right)^m$ for $m=0,1,...,n$. Each term in this trace involves $2km$ copies of the original system. Applying the replica wormhole calculation to this computation, it is easy to convince ourselves that replica wormhole will only appear between $\rho_A$ and $Z_A^{(k)}$ within each trace. We can label the $2km$ replica by $(a,s)$ with $a=1,2,...,m$ and $s=1,2,...,2k$, with $(a,1)$ the $m$ copies of $\rho_A$, and $(a,s),~s\geq 2$ the $2k-1$ copies of $\rho_A$ or $\sigma_A$ in the operator $Z_A^{(k)}$. The replica wormhole will only connect $(a,s)$ with $(a,s+1)$ (with $s=2k+1$ identified with $s=1$). This is illustrated in Fig. \ref{fig:replica Zk}. Compare this calculation with the ``ordinary'' case, such as the calculation of ${\rm tr}(\rho_A^n)$, we see that the dominant saddle point configuration contains wormholes connecting the ``physical geometry'' with ``simulated geometry'' in $Z_A^{(k)}$. This is the possibility that was missed in the derivation of the one norm correspondence in Eq. (\ref{eq:one norm}). 


In summary, what we learned from miracle operators is that carrying a particular measurement can not only modify the bulk geometry of the black hole, but even modify the boundary condition of the geometry. Instead of being a smooth geometry with boundary condition set by the flat space region, the geometry now contains wormholes that connect the system with the operator being measured. On comparison, if we view ${\rm tr}(\rho_A^n)$ as an expectation value of cyclic permutation operator $X_A$ in $n$ copies of the system, then this operator induces replica wormhole between different replicas, but does not change the boundary condition of the geometry. If we take the point of view that gravity is an ensemble theory\cite{saad2019jt,stanford2019jt,Saad:2019pqd,bousso2020gravity,marolf2020transcending,Stanford:2020wkf,Pollack:2020gfa,Afkhami-Jeddi:2020ezh,Maloney:2020nni,Belin:2020hea,Cotler:2020ugk,Chen:2020ojn, Hsin:2020mfa,Belin:2020jxr, Milekhin:2021lmq}, then we believe that miracle operators differ from ordinary operators by ``knowing" the random parameters in the gravity theory.\cite{qi2021holevo} For example, if we are considering a Sachdev-Ye-Kitaev (SYK) model\cite{sachdev1993gapless,kitaev2014hidden,kitaev2015simple} coupled with bath, the miracle operators have to depend on the random coupling $J$ in the SYK model. In more general cases, the natural of the random parameters remain an open question, but results in random tensor network models\cite{hayden2016holographic} suggest that such random parameters exist not only in black hole systems but also in spacetime geometry without black hole, such as the AdS vacuum\cite{qi2021holevo}. 


Another point we want to highlight is that such a dramatic effect to spacetime geometry occurs even if we are only carrying a binary measurement that learns about a single bit of information in the interior.

\section{The firewall}\label{sec:firewall}

We would like to investigate the implication of our result in understanding the firewall paradox\cite{almheiri2013black,almheiri2013apologia}. The firewall paradox points out that after Page time, a Hawking radiation mode $b$ is entangled with the earlier radiation, so that if it is also entangled with its partner behind the horizon $\tilde{b}$, this will violate strong subadditivity (SSA) of entanglement entropy. The entanglement island result points out that the SSA argument does not apply, since $\tilde{b}$ in the island is not independent from the earlier radiation. In Fig. \ref{fig:firewall1} (a), if $\tilde{b}$ is in the entanglement island of $A$ region, and $b$ is outside $A$ region, then it is ok to have $b$ entangled with $\tilde{b}$ and also entangled with $A$, since $\tilde{b}$ is actually part of $A$. However, we would like to address more concretely whether an infalling observer will see a firewall. We will study two different setups, illustrated in Fig. \ref{fig:firewall1} (a) and Fig. \ref{fig:entanglement check} (a). In both cases there is an EPR pair of modes $b,\tilde{b}$ and $\tilde{b}$ is in the entanglement island of a radiation region $A$. The difference is that in the first setup the partner $b$ is outside region $A$ while in the second setup it's in $A$. In both cases, an infalling observer $X$ carries a Bell basis measurement to detect if $b\tilde{b}$ is in one of the maximally entangled states. The first setup addresses whether a measurement on $A$ can act on $\tilde{b}$ and destroy the entanglement between $b$ and $\tilde{b}$, {\it i.e.} creates a firewall. The second setup addresses whether an entanglement checking experiment can be applied to $A$, now including the information of both $b$ and $\tilde{b}$, and whether such an entanglement checking will create a firewall seen by the infalling observer $X$. In the following we will discuss these two setups in two subsections.

\subsection{Entanglement breaking measurement}

To be concrete, let us assume $b$ and $\tilde{b}$ are local wavepackets of bosons, such as photons. Denote $\hat{b}^\dagger_\sigma$ and $\hat{\tilde{b}}^\dagger_\sigma$ as the creation operators of these two modes respectively, with $\sigma=\uparrow,\downarrow$ two states of spin or other internal states. $\hat{b}^\dagger,\hat{\tilde{b}}^\dagger$ are defined in the QFT Hilbert space. Denote the state without this pair as $\rhoqft$, then we can create an EPR pair in one of the Bell basis states:
\begin{align}
    \sigmaqft&=\hat{\Delta}_\pm^\dagger\rhoqft\hat{\Delta}_\pm\nonumber\\
    \hat{\Delta}_\pm&=\frac1{\sqrt{2}}\left(\hat{b}_{\uparrow}\hat{\tilde{b}}_{\downarrow}\pm\hat{b}_{\uparrow}\hat{\tilde{b}}_{\downarrow}\right)\label{eq:sigmapm}
\end{align}
In this subsection we only need one of them, which we will pick as $\sigmaqft_-$. In next subsection we will need both. 

\begin{figure}
    \centering
    \includegraphics[width=5.3in]{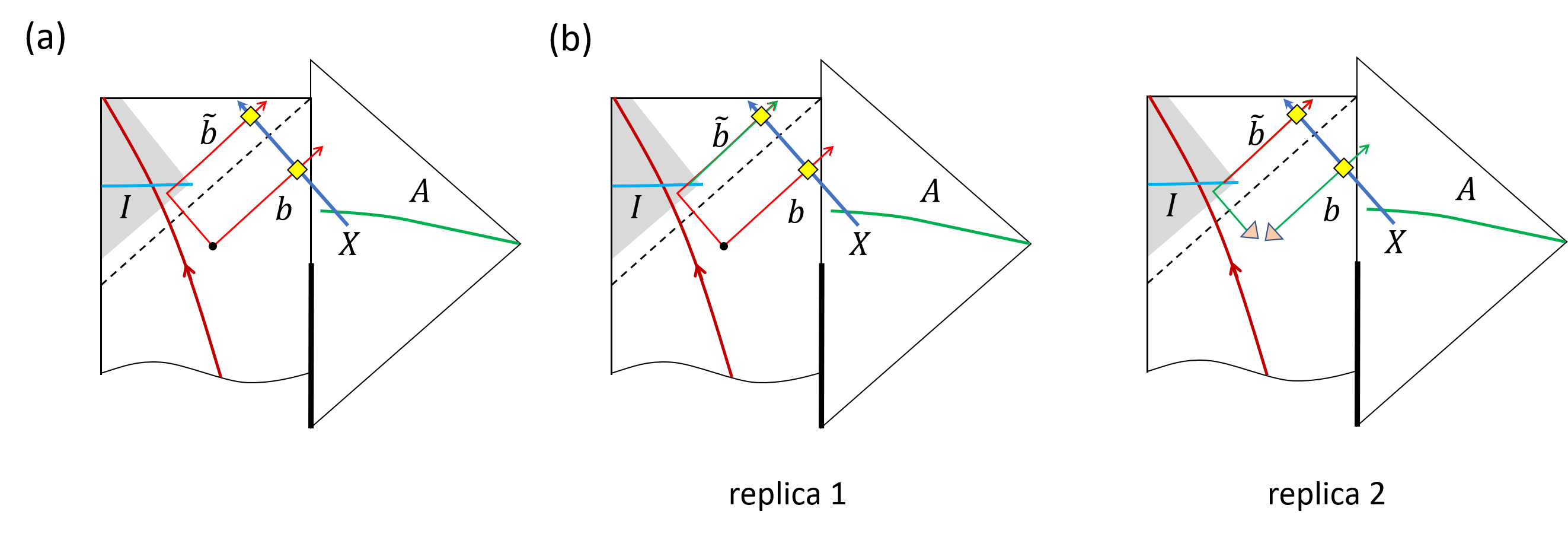}
    \caption{(a) Illustration of a pair of qubits $b,\tilde{b}$, when $\tilde{b}\in I$ and $b\notin A$. An infalling observer $X$ carries a binary measurement that checks if $b\tilde{b}$ is in the particular entangled state. (b) Illustration of the replica geometry that applies operator $P_A^{\uparrow \downarrow}$ which distinguishes the two states with opposite spin of $\tilde{b}$. The green and red lines represent the trajectory of $b$ and $\tilde{b}$ in the two replicas respectively. (In general there are $n$ replicas. We draw $n=2$ here for simplicity.) Since only one of them crosses the branchcut line, $X$ will not see the partner $\tilde{b}$ but see the copy of $\tilde{b}$ in a different replica, leading to failure of entanglement checking.}
    \label{fig:firewall1}
\end{figure}

Now when the system is in state $\sigmaqft_{-}$, an infalling observer $X$ brings $b$ into the interior and check whether $b\tilde{b}$ is in the particular entangled state created by $\hat{\Delta}_-$. The measurement operator is
\begin{align}
    P_X=\hat{\Delta}_-^\dagger \hat{\Delta}_-\label{eq:infalling measurement}
\end{align}
Without other intervention to the system, $P_X$ will return eigenvalue $1$ with probability $1$. 
$X$ records the result of the measurement. 

Now we would like to carry a measurement on $A$ that measures the spin $z$ component of $\tilde{b}$. For that purpose let us consider two other states
\begin{align}
    \sigmaqft_{\uparrow}=\hat{\tilde{b}}_\uparrow^\dagger \hat{b}_\downarrow^\dagger\rhoqft \hat{b}_\downarrow \hat{\tilde{b}}_\uparrow,~ \sigmaqft_{\downarrow}=\hat{\tilde{b}}_\downarrow^\dagger \hat{b}_\uparrow^\dagger\rhoqft \hat{b}_\uparrow\hat{\tilde{b}}_\downarrow
\end{align}
which contains $Z$ eigenstates of $\tilde{b}$ qubit. Then we can define the optimal projection operator on $A$ defined in Eq. (\ref{eq:operator replica}) that distinguishes these two states:
\begin{align}
    \hat{Z}_A&\equiv 2P_A^{\uparrow \downarrow}-\mathbb{I}_A={\rm sgn}\left(\sigma_{\uparrow,A}-\sigma_{\downarrow,A}\right)
\end{align}
We implement this operator by the replica trick in Eq. (\ref{eq:operator replica}):
\begin{align}
    {\rm tr}\left(\sigma_{-,A}\hat{Z}_A\right)&={\rm tr}\left.\left[\sigma_{-,A}\left(\sigma_{\uparrow, A}-\sigma_{\downarrow,A}\right)^{2n-1}\right]\right|_{n\rightarrow \frac12}
\end{align}
Every term in the righthand side of this equation involves $2n$ replicas of the original system, with a branchcut at $A$ in the same way as in Renyi $2n$ entropy calculation. In the limit when the back reaction caused by the modes $\tilde{b},b$ is negligible, the $2n$ replica geometry contains the replica wormhole in the same way as in the Renyi entropy calculation, as is illustrated in Fig. \ref{fig:firewall1} (b). Therefore in the saddle point geometry, there is a branchcut at $A$ and $I$, where different replicas are connected by cyclic permutation. In the first replica there is an entangled pair of modes, while in the other $2n-1$ replicas there is a pure state of $\tilde{b}b$. One should remember that the infalling observer $X$ is also part of the system, so it should also be present in each replica.

Now it is important to remember that the infalling observer is also part of the system, which should appear in each replica. The question is whether the entanglement checking measurement $P_X$ still succeeds in this replicated geometry. As is illustrated in Fig. \ref{fig:firewall1} (b), $\tilde{b}$ goes across the branchcut line in $I$, but $b$ does not, so that when the infalling observer in $k$-th replica brings $b$ in that replica to the interior, it meets with the $\tilde{b}$ mode in the $k+1$-th replica. Consequently, none of the infalling observers will see an entangled pair of $b\tilde{b}$. In other words, the entanglement checking fails for all replica calculation with $2n$ replicas. The observer that was originally in replica $1$ (the one with an entangled pair) will see a state
\begin{align}
    \rho_{b\tilde{b}}=\frac12\mathbb{I}_b\otimes\left(\ket{\sigma}\bra{\sigma}\right)_{\tilde{b}},~\sigma=\uparrow\text{~or~}\downarrow
\end{align}
such that
\begin{align}
    \langle \hat{P}_X\rangle=\frac14
\end{align}

Since this result applies to all replica number $2n$, it is reasonable (although not rigorous) to suggest that the same holds when taking analytic continuation $n\rightarrow 1/2$, which means the infalling observer will see the $b\tilde{b}$ entanglement destroyed. In this sense the measurement $P_A^{\uparrow \downarrow}$ creates a firewall at the quantum extremal surface, although it is not a firewall in the sense of energy. Degrees of freedom across the quantum extremal surface, as are seen by the infalling observer, are not entangled because they are entangled with partners in another replica, but this does not require a high energy excitation.

It should be noted that this firewall creation occurs whether or not $X$ jumps in before or after the measurement on $A$ is carried. This looks nonlocal but does not violate causality, because the action we take on $A$ is a measurement. In ordinary quantum systems, it is also allowed that two space-like separated measurements have correlated results. The calculation above shows that there is a firewall {\it condition} on a given output of the measurement $P_A^{\uparrow \downarrow}$. 

\subsection{Entanglement checking}

\begin{figure}
    \centering
    \includegraphics[width=5.3in]{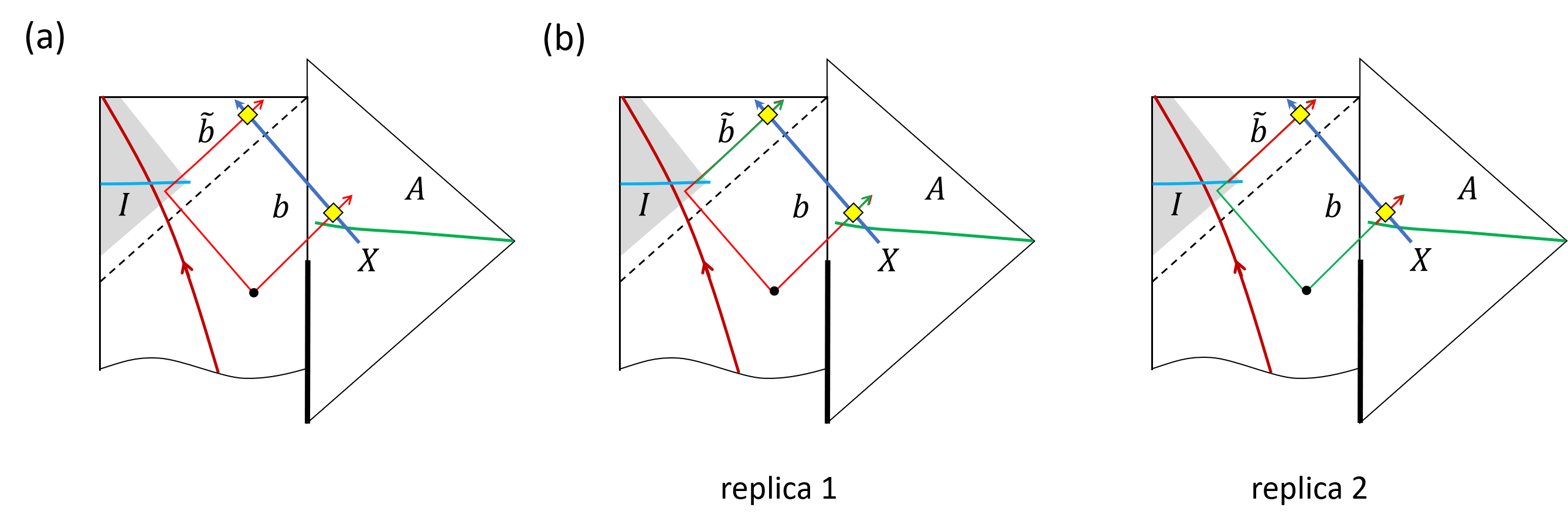}
    \caption{(a) Illustration of a pair of qubits $b,\tilde{b}$ with $\tilde{b}\in I,~b\in A$. An infalling observer $X$ measures whether $\ket{b\tilde{b}}$ is in a particular maximally entangled states such as $\ket{-}$. (b) Illustration of the replica geometry that applies operator $P_{A}^{+-}$ (defined in Eq. (\ref{eq:DefPApm})) which distinguishes the two states $\ket{+},\ket{-}$ in $A$. The green and red lines represent the trajectory of $b$ and $\tilde{b}$ in the two replicas respectively, in the same way as in Fig. \ref{fig:firewall1}. We see that the entanglement checking of $X$ will still succeed because both $b$ and $\tilde{b}$ are acted by the same cyclic permutation operation.}
    \label{fig:entanglement check}
\end{figure}

Instead of measuring the state of $\tilde{b}$, we would like to consider an entanglement checking measurement applied to the exterior. In the setup of Fig. \ref{fig:entanglement check} (a), we consider a situation when $b$ has already entered region $A$. $\tilde{b}$ is in the entanglement island of $A$, so that it should be possible to check the entanglement of $b\tilde{b}$ in region $A$. For example we can apply the Petz map to the operator $P_X$ applied by the infalling observer. Instead of Petz map, we prefer to make use of the optimal projector construction. For that purpose we consider two states $\sigma_{\pm,QFT}$ with two orthogonal maximally entangled states of $b\tilde{b}$, defined in Eq. (\ref{eq:sigmapm}). Then we can define the projection operator $P_A^{+-}$ that distinguishes these two states
\begin{align}
    \hat{F}_A&\equiv 2P_A^{+-}-\mathbb{I}_A={\rm sgn}\left(\sigma_{+,A}-\sigma_{-,A}\right)\label{eq:DefPApm}
\end{align}
Now if we take the state $\sigma_{-}$ and measure $\hat{F}_A$, by construction we obtain
\begin{align}
    {\rm tr}\left(\sigma_{-,A}\hat{F}_A\right)=\left.{\rm tr}\left[\sigma_{-,A}\left(\sigma_{+,A}-\sigma_{-,A}\right)^{2n-1}\right]\right|_{n\rightarrow \frac12}=-1\label{eq:entanglement checking}
\end{align}
This is because one term in the expansion $-{\rm tr}\left(\sigma_{-,A}^{2n}\right)\rightarrow -1$ when $n\rightarrow \frac12$, and all other terms are of the form
\begin{align}
    {\rm tr}\left(\sigma_{+,A}^a\sigma_{-,A}^b...\right)\simeq e^{-\mathcal{A}_{grav}}{\rm tr}\left(\sigmaqft_{+,AI}^{a}\sigmaqft_{-,AI}^b...\right)
\end{align}
which contains a products of $\sigmaqft_{+,AI}$ and $\sigmaqft_{-,AI}$. Since $\sigmaqft_{+}$ and $\sigmaqft_{-}$ contains two orthogonal states of the $b\tilde{b}$ pair, we have $\sigmaqft_{+,AI}\sigmaqft_{-,AI}=0$, so that these terms all vanish. (The gravity contribution $\mathcal{A}_{grav}\simeq (2n-1)\frac{\left|\partial I\right|}{4G_N}$ in the $n\rightarrow \frac12$ limit.)

In parallel with the discussion in previous subsection, we can ask what happens to infalling observer $X$ which carries the measurement $\hat{P}_X$ in Eq. (\ref{eq:infalling measurement}). Different from the previous setup, now $b$ goes through the branchcut at $A$ and $\tilde{b}$ goes through that at $I$, so that they can still meet in the same replica. The observer $X$ in each replica will observe a maximal entangled state $\ket{-}$ since the only nonvanishing term contains $\sigma_{-,A}^{2n}$. Therefore we would argue that in the analytic continuation to $n\rightarrow \frac12$, the entanglement checking experiment by observer $X$ is successful with probability $1$. It is interesting to note that the trajectory of $b$ and $\tilde{b}$ together form a noncontractable loop in the replica manifold ({\it i.e.} a nontrivial generator of the first homotopy group). The difference between the two setups with $b$ outside $A$ (Fig. \ref{fig:firewall1}) and $b$ inside $A$ (Fig. \ref{fig:entanglement check}) is topological.

In short, we see that both the exterior and infalling entanglement checking can succeed, at least in this particular choice of measurements. Roughly speaking, this is possible because the measurement in $A$ has to create other copies of the universe, including the infalling observer. Another implicit assumption (which seems physically reasonable) is that the infalling observer is defined for each replica geometry separately, and there is no ``super-observer'' that can take different replicas and apply a joint measurement. Our result seems to be consistent with the result of A. Almheiri\cite{almheiri2021comments} in the final state projection model\cite{horowitz2004black}. The gravitational calculation clarifies that entanglement checking experiment in $A$ does not have to apply a projection to many qubits, but can extract only one bit of information about whether $b\tilde{b}$ is in the particular entangled state $\ket{-}$. We realize that the argument about analytic continuation in this section is not rigorous. More careful analysis is a task for future work.

\section{Further discussion and conclusion}\label{sec:conclusion}

In summary, we have shown that there is a paradox if we assume the QES formula applies to an AdS black hole coupled with a radiation system with arbitrary dynamics. When the geometry is semiclassical, and the QES formula applies, information about the interior is not available to any observer who is only coupled with the radiation through LOCC (and does not have quantum entanglement with the radiation and black hole to start with). On the other hand, information in the entanglement island of a region $A$ can be reconstructed in $A$, which suggests that there must exist miracle operators which induce qualitative change of bulk geometry when they are applied to $A$, even if only one bit of information is obtained in this measurement. We have explicitly constructed such an operator which distinguishes two given states optimally, and show that in a replica calculation, the large geometry fluctuation corresponds to replica wormhole connecting the original universe with new copies of the universe contained in the projection operator itself. This construction helps us to address some questions in the firewall paradox. We show that a projective measurement in a radiation region $A$ can break the entanglement between two modes across the QES surface, {\it i.e.} the boundary of the entanglement island. On the other hand, an entanglement checking measurement that verifies the entanglement between a Hawking qubit $b$ and the earlier radiation does not affect the entanglement checking measurement of an infalling observer, at least in the particular setup we consider. 

There are many open questions related to this setup. One question is whether the difference between regular operators and miracle operators is a signature of gravity theory being an ensemble theory.\cite{maldacena2004wormholes,saad2019jt,stanford2019jt,penington2019replica,marolf2020transcending} If gravity is an effective description of an ensemble average over a family of boundary theories, then it is natural to distinguish between operators that does and does not depend on random parameters in the boundary theory. Only the latter will possibly detect the interior. A nontrivial question is whether wormholes connecting physical systems with simulated systems are still well-defined if there is no ensemble averaging.

Another question is whether miracle operators can still be defined when gravity in the bath is dynamical. For example, in a flat space Schwarzchild black hole, can one define miracle operators that create simulated copies of spacetime and replica wormholes? Ref.\cite{dong2020effective} proposed a generalization of the entropy calculation of a bath region to the dynamical gravity case, but in that generalization the entropy is a measure of ``effective" uncertainty in the state of low energy degrees of freedom. Roughly speaking, we can view the effective entropy as the entropy of a ``conditional state" of low energy degrees of freedom, with the condition that a classical background geometry is observed by a family of observers in the bath region. It is unclear to the author how this discussion will be modified if we more rigorously taken into account the nonlocality of quantum gravity Hilbert space\cite{geng2021inconsistency,geng2021information,chowdhury2021holography,chowdhury2021physical}. For example, we notice that Ref.\cite{geng2021inconsistency} proposed that entanglement islands can only be rigorously defined in systems with massive gravitons. 

Ref. \cite{bousso2014measurements} pointed out that in the final state projection proposal\cite{horowitz2004black} there is a problem with probability interpretation of the measurement carried by an infalling observer, because of the failure of decoherence between different histories. We have not addressed this problem in our discussion of infalling observer. For example consider an observer $X$ made of a large number of spins with the initial state $\left(\ket{\uparrow}\bra{\uparrow}\right)^{(\otimes M)}$. The measurement process is a unitary operator 
\begin{align}
    U_M=P_X\otimes \mathbb{I}_X+\left(\mathbb{I}-P_X\right)\otimes X_X
\end{align}
with $X_X$ flips all the spins in $X$. In ordinary quantum measurement, we apply $U_M$ to obtain $U_M\rho_{b\tilde{b}}\otimes\left(\ket{\uparrow}\bra{\uparrow}\right)^{(\otimes M)}U_M^\dagger$ and then trace over any one of the $M$ qubits in $X$. This removes the off-diagonal terms and leads to a measurement channel
\begin{align}
    \mathcal{C}_M\left(\rho_{b\tilde{b}}\otimes\left(\ket{\uparrow}\bra{\uparrow}\right)^{(\otimes M)}\right)=P_X\rho_{b\tilde{b}}P_X\otimes \left(\ket{\uparrow}\bra{\uparrow}\right)^{\otimes (M-1)}+\left(\mathbb{I}-P_X\right)\rho_{b\tilde{b}}\left(\mathbb{I}-P_X\right)\otimes \left(\ket{\downarrow}\bra{\downarrow}\right)^{\otimes (M-1)}
\end{align}
When $X$ is behind the horizon, it is unclear whether we are still allowed to do the partial trace. On the other hand, considering that horizon is not a local concept, it is possible that we are crossing the horizon of a giant black hole now, which should not affect the probabilistic interpretation of the quantum experiments that occur in a physics lab at this moment. This suggests that we should be able to talk about what an infalling observer sees, in the same way as an outside observer. A more rigorous formalism is a task for future research.

Another question is how rare are the set of miracle operators. We know the special constructions discussed here. It will be interesting to have a more precise statement about how the miracle operators are very rare. In the entanglement checking setup in Fig. \ref{fig:entanglement check}, if we define the optimal operator for two states $\sigma_{1}$ and $\sigma_{2}$ which are not orthogonal, all terms in Eq. (\ref{eq:entanglement checking}) are nonzero. This creates a linear superposition of different ``branches" in which infalling observer $X$ can observe different results. This is not immediately contradictory because the two states are not entirely distinguishable so the observer always has a chance to mistaken state $1$ as state $2$. However, it requires a more thorough investigation to understand whether all observations of the infalling observer are consistent with quantum mechanics. This question seems to be related to the discussion about state-dependence of the reconstruction map\cite{papadodimas2013infalling,hayden2019learning}. 

It is interesting to think this setup as an example of quantum algorithmic measurement (QUALM)\cite{aharonov2021quantum}. A QUALM is a quantum algorithm that contains a known part controlled by the experimentalist, and an partially unknown part controlled by nature that contains some hidden parameters. The purpose of the algorithm is to find out some message about the unknown parameters. In the black hole problem, as is illustrated in Fig. \ref{fig:locc}, the known part is the coupling between ancilla $W$ and the radiation $R$, and the dynamics of $R$ itself. The partially unknown part is the black hole $B$ and its coupling with radiation $R$. In the QUALM language, what we have shown is that the task of finding out the information in an interior qubit is very difficult for an observer with incoherent access to the system, if the observer does not know the black hole microstate. It is reasonable to guess that the difficulty ({\it i.e.} QUALM complexity of the task) is exponential in $1/G_N$, since a saddle point with nontrivial island only has a contribution to the partition function that is exponential in $1/G_N$.  It is interesting to relate this result to that of Ref. \cite{aharonov2021quantum} and see whether it is possible to make our result more rigorous in the QUALM framework.

\noindent{\bf Acknowledgement.} We would like to thank Ahmed Almheiri, Yiming Chen, Xi Dong, Daniel Harlow, Hong Liu, Donald Marolf, Daniel Ranard, Douglas Stanford, Zhenbin Yang and Pengfei Zhang for helpful discussions. This work is supported by the Simons Foundation, and in part by the DOE Office of Science, Office of High Energy Physics, the grant de-sc0019380. This work was developed in part at the Kavli Institute for Theoretical Physics during the workshop ``Gravitational Holography", which is supported in part by the National Science Foundation under Grant No.\ PHY-1748958.

\appendix  

\section{Definition of LOCC}\label{app:LOCC}

In this appendix we give a precise definition of LOCC. For a useful reference, see Ref. \cite{chitambar2014everything}. (In general LOCC is defined for multiple parties, but here we will only discuss the two party case.) For two quantum systems $A$ and $W$, with the Hilbert space $\mathbb{H}_A\otimes \mathbb{H}_W$, a one-way LOCC from $A$ to $W$ is a quantum channel of the following form:
\begin{align}
    \mathcal{C}_{A\rightarrow W}=\sum_{n=1}^M\mathcal{M}_n^A\otimes \mathcal{N}_n^W
\end{align}
Here $\mathcal{M}_n^A$ are completely positive (CP) maps with the additional condition that  $\sum_n\mathcal{M}_n^S$ is a completely positive and trace-preserving (CPTP) map. In other words, for any density operator $\rho$ of subsystem $A$, $\sum_n\mathcal{M}_n^S(\rho)$ is a density operator with unit trace. $\mathcal{N}_n^W$ for each $n$ is a CPTP map. Physically, if we apply $\mathcal{C}_{A\rightarrow W}$ to a product state $\rho_A\otimes \sigma_W$, we obtain
\begin{align}
    \mathcal{C}_{A\rightarrow W}\left(\rho_A\otimes \sigma_W\right)=\sum_{n=1}^M\mathcal{M}_n^A\left(\rho_A\right)\otimes \mathcal{N}_n^W\left(\sigma_W\right)
\end{align}
which is a separable state. Define
\begin{align}
    p_n&={\rm tr}\left(\mathcal{M}_n^A\left(\rho_A\right)\right)\\
    \tilde{\rho}_{An}&=p_n^{-1}\mathcal{M}_n^A\left(\rho_A\right)
\end{align}
the channel $\mathcal{C}_{A\rightarrow W}$ applies a weak measurement to $A$ and if the measurement output is $n$, apply channel $\mathcal{N}_n^W$ to $W$. 

A $r$-round LOCC between $A$ and $W$ contains $r$-round of back-and-forth communication between these two systems:
\begin{align}
    \mathcal{C}_{AW}^{(r)}=\mathcal{C}_{W\rightarrow A}^{r}\circ\mathcal{C}_{A\rightarrow W}^{r}\circ \mathcal{C}_{W\rightarrow A}^{r-1}\circ\mathcal{C}_{A\rightarrow W}^{r-1}\circ...\circ\mathcal{C}_{W\rightarrow A}^{1}\circ\mathcal{C}_{A\rightarrow W}^{1}
\end{align}

\section{Positive conditional entropy for separable states}\label{app:conditional entropy}

For completeness, we include a proof that the conditional entropy is always positive for separable states.

For a separable state
\begin{align}
    \rho_{WI}=\sum_{i}p_i\rho_{Wi}\otimes \rho_{Ii}
\end{align}
we consider an auxiliary state
\begin{align}
    \pi_{\tilde{W}I}=\sum_ip_i\ket{i}\bra{i}\otimes \rho_{Ii}
\end{align}
with $\ket{i}$ an orthonormal basis in the $\tilde{W}$ system. The mutual information between $\tilde{W}$ and $I$ is 
\begin{align}
    I(\tilde{W}:I)=S\left(\sum_ip_i\rho_{Ii}\right)-\sum_ip_iS\left(\rho_{Ii}\right)
\end{align}
Which is also the Holevo information that measures the amount of information that can be read out from $I$ about the classical message in $\tilde{W}$. Therefore
\begin{align}
    I\left(\tilde{W}:I\right)_\pi\leq S_{I}\equiv S\left(\sum_ip_i\rho_{Ii}\right)
\end{align}
Now we can apply a measure-and-prepare channel to map a state in $\tilde{W}$ to that in $W$:
\begin{align}
    \mathcal{C}\left(\sigma_{\tilde{W}}\right)=\sum_i\bra{i}\sigma_{\tilde{W}}\ket{i}\rho_{Wi}
\end{align}
The channel maps the auxiliary state $\pi_{\tilde{W}I}$ to $\rho_{WI}$. Since mutual information cannot be increased by a quantum channel, we obtain that in $\rho_{WI}$
\begin{align}
    I(W:I)_\rho\leq I\left(\tilde{W}:I\right)_\pi\leq S_{I}
\end{align}
Therefore
\begin{align}
    S(W)\leq S(WI)
\end{align}

\bibliographystyle{jhep}
\bibliography{refs.bib,refs2.bib}

\end{document}